\documentclass[9pt,twocolumn,twoside]{opticajnl}
\journal{opticajournal} 
\setboolean{shortarticle}{false}

\usepackage{siunitx}
\usepackage{hyperref}
\usepackage{comment}
\usepackage{graphicx}
\usepackage{amsmath}
\usepackage{amssymb}
\usepackage{bm}

\usepackage{xcolor}

\usepackage{physics}

\title{Real-time estimation of the transmission matrix of an atmospheric channel}
\author[1]{Cade Peters}
\author[2]{Raphael Bellossi}
\author[2]{Douglas McDonald}
\author[1]{Andrew Forbes}
\author[2]{Szymon Gladysz}
\author[2,*]{Giacomo Sorelli}

\affil[1]{School of Physics, University of the Witwatersrand, Private Bag 3, Wits 2050, South Africa}

\affil[2]{Fraunhofer IOSB, Ettlingen, Fraunhofer Institute of Optronics, System Technologies and Image Exploitation, Gutleuthausstr. 1, 76275 Ettlingen, Germany}

\affil[*]{giacomo.sorelli@iosb.fraunhofer.de}

\begin{abstract}  
Optical wavelengths have received significant attention in free-space channels and are vital for applications in communications, imaging and sensing. Their widespread implementation is motivated by a variety of factors including improved bandwidths, increased security and higher energy efficiency. 
However, these advantages cannot be fully brought to bear in real-world scenarios due to the deleterious effects of atmospheric turbulence. 
Induced by small temperature and pressure fluctuations in the environment that vary rapidly in space and time, these effects cause significant power losses which decrease SNR, induce severe crosstalk in communication links, and greatly limit resolution of long-range imaging systems. 
To overcome this, we numerically and experimentally investigate the reconstruction of the transmission matrix of a time-evolving atmospheric channel with a real-time recursive optimization routine.
We demonstrate that this estimation technique is able to keep up with the evolution of the channel and enables a significant improvement of communication-relevant quantities such as the coupling of the received light into a single-mode fiber while notably reducing the probability and duration of power outages, even in strong turbulence. 
Our results have immediate applications in free-space optical communication in both the classical and quantum regimes.
\end{abstract}

\begin{document}

\maketitle
\section{Introduction}
Atmospheric channels play an important role in various fields of science and technology including astronomy, remote sensing, imaging, and communication \cite{Roddier04, Andrews05}.
All these rather diverse applications share the need of efficiently transmitting and/or receiving information that has been encoded into electromagnetic radiation traveling through the channel. In recent years, optical wavelengths, and in particular free-space optical (FSO) links \cite{trichili2019communicating,trichili2020roadmap}, have received significant attention driven by the need for higher communication bandwidths \cite{Richardson1,richardson2013space}, increased security \cite{lopez2015physical}, lower energy consumption \cite{miller2017attojoule}, and a means to bridge the digital divide in a manner that is cost effective and license-free \cite{lavery2018tackling}.

However, atmospheric channels exhibit rapidly evolving fluctuations in their refractive index profile caused by changes in temperature and pressure. These lead to noticeable phase fluctuations in the near-field and scintillation and significant phase perturbations upon propagation and in the far-field \cite{peters2025structured}. These distortions degrade the system's point-spread function \cite{racine1996telescope}, cause modal scattering which leads to significant crosstalk in mode-division multiplexing schemes \cite{cox2020structured}, and induce significant power losses, all of which increase bit-error rates in communication \cite{Sorelli19, Cox19, Gu20} while decreasing resolution when imaging \cite{yaqoob2008optical,ding2024wavefront}. Accordingly, it is crucial to identify strategies which efficiently mitigate the random wave distortions induced by the atmosphere.

For the past few decades, the main turbulence compensation strategy has been adaptive optics (AO) which uses a fast wavefront sensor working in closed loop with a wavefront shaping element (generally a deformable mirror) to correct turbulence-induced phase distortions in real-time \cite{Tyson16} and mitigate some of the channels deleterious effects \cite{Gehner20, Anzuola16, Yu17, Abado10}.
Correcting for only phase fluctuations is sufficient for weak atmospheric layers, especially if located close to the receiver, i.e., in astronomy. Critically, AO fails to compensate for intensity fluctuations which become important in horizontal atmospheric channels, particularly under strong turbulence conditions \cite{Botygina_2019, Lukin_2020}, which can have a significant impact, e.g., in free-space quantum or classical communication \cite{Zhao20}.

In contrast, studies involving \textit{static} scattering media have leveraged advanced wavefront shaping techniques, that have quickly become the preferred tool to control light propagation through biological and other similar complex channels \cite{Vellekoop:15}.
This control is achieved by measuring the transmission matrix (TM) of the channel \cite{Popoff10} and has enabled numerous advances including improved focusing \cite{Vellekoop07}, point-spread function engineering \cite{Boniface17}, high energy transmission (even through multiple scattering media) \cite{Kim12}, and arbitrary field transmission \cite{Devaud22}. This success has prompted the use of TMs for the mitigation of atmospheric turbulence, with recent works showing it is possible to determine invariant states of a turbulent channel \cite{peters2025tailoring}, even for thick, horizontal paths \cite{Klug23} and study the temporal evolution of its highly transmitting modes \cite{Bachmann23}. While promising, these works assumed perfect knowledge of the channel's TM. Unlike typical scattering media, where the channel evolves relatively slowly and can be measured or estimated within reasonable time-constraints \cite{Valzania23}, atmospheric variations are of the order of a few milliseconds \cite{fried1990greenwood} making it extremely difficult to measure the full TM of the channel due to the limited refresh rates of conventional hardware.



In this work, we investigate the feasibility of assessing the time-dependent TM of a turbulent channel in real-time by employing a recursive algorithm for its real-time estimation, while emulating realistic bandwidths of state-of-the-art wavefront shaping and sensing devices, such as those employed in high-end AO systems. 
We show, by thorough experimental demonstrations and extensive wave-optics simulations, that this approach is suitable for the optimization of communication-relevant performance metrics, such as the total power coupled into a single-mode fiber, enabling a significant performance increase compared to transmission of standard modal bases, especially in strong turbulence.  

\section{Methods}
\label{sec:methods}
\subsection{Modeling a time-dependent atmospheric channel}
\label{sec:Model}
\begin{figure}
    \centering
    \includegraphics[width = \linewidth]{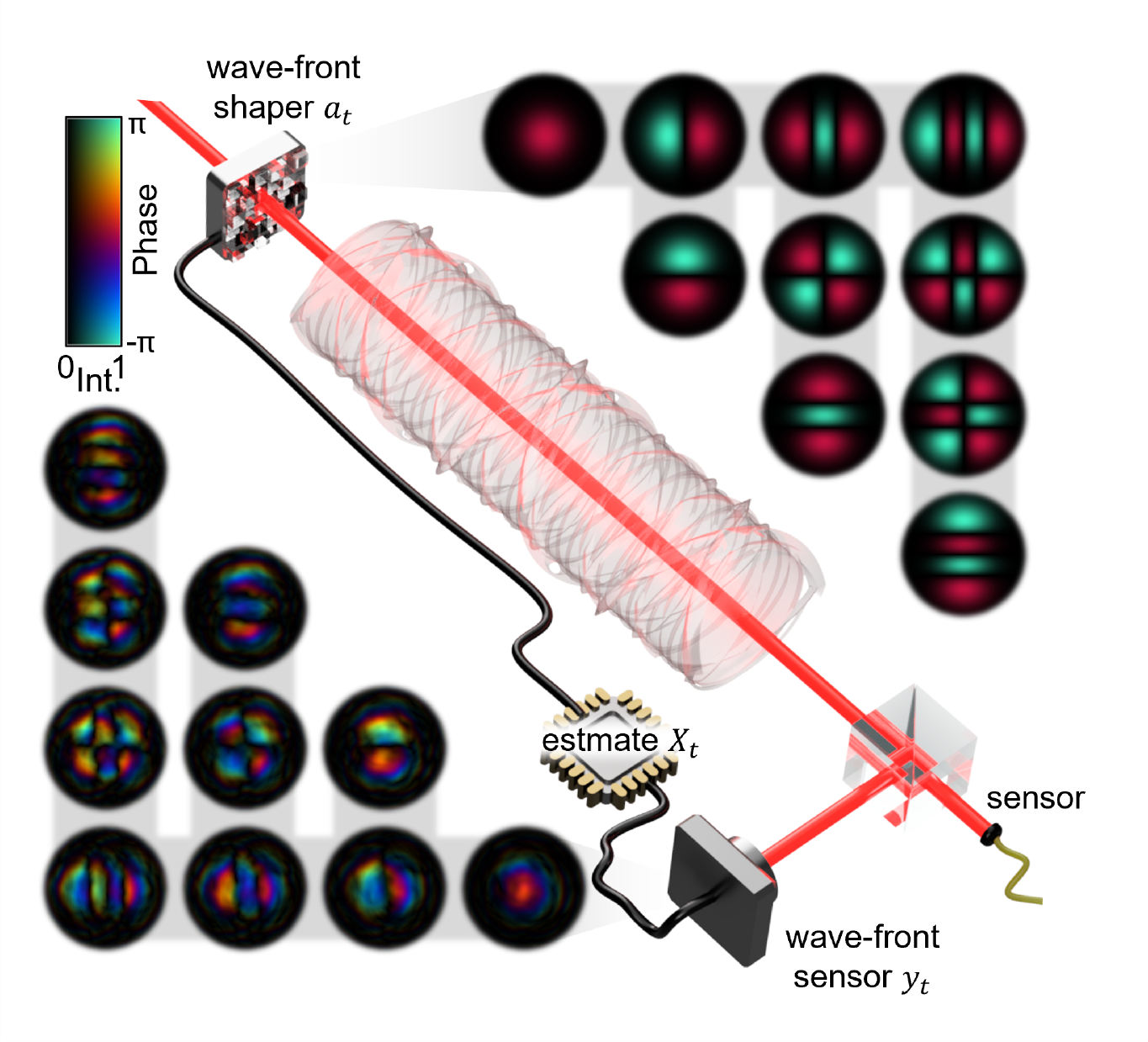}
    \caption{Schematic illustration of the concept for the real-time estimation of an atmospheric channel: a series of probe modes $\vec{a}_t$ (here Hermite-Gaussian modes) are successively transmitted through the channel and measured at the output as $\vec{y}_t$. The input-output pairs $\vec{a}_t$ and $\vec{y}_t$ are used to update the estimation of the dynamic transmission matrix $X_t$, whose knowledge is used to compute the input field that optimizes a given performance metric, such as the power coupled into a single-mode fiber. }
    \label{fig:scheme}
\end{figure}
Propagation of a monochromatic wave $u(\bf{r})$ with wavelength $\lambda$ through an atmospheric channel is described by the {\it stochastic parabolic equation} \cite{Andrews05}
\begin{equation}
    - 2ik\,\frac{\partial u (\bf{r})}{\partial z} =  \nabla^2_\perp u({\bf r}) + 2k^2\,\delta n({\bf r}, t)\, u({\bf r}),
    \label{eq:stoch_par}
\end{equation}
where $k = 2\pi/\lambda$ is the wave number, $\nabla^2_\perp$ is the transverse (with respect to the propagation direction $z$) Laplace operator, and $\delta n({\bf r}, t)$ describes the turbulence-induced spatial and temporal fluctuations of the refractive index of the channel. 
The statistics of such random fluctuations are captured by the two-point correlation function, $\langle \delta n({\bf r}, t), \delta n({\bf r}^\prime, t) \rangle$, whose Fourier transform $\Phi_n(\kappa)$ is known as the {\it refractive index power spectrum}.
According to the Kolmogorov theory of turbulence \cite{Kolmogorov41a}, the three-dimensional refractive index power spectrum is given by
\begin{align}
    \Phi_n(\kappa) =  0.033\, C_n^2\, \kappa^{-11/3}, 
    \label{eq:kol_spec}
\end{align}
where $\kappa = |\bm {\kappa}|$, $\bm {\kappa}$ is the transverse angular wavenumber, and $C_n^2$ is the refractive index structure constant. 

For a wave propagating a distance $L$ through an atmospheric channel, the random spatial inhomogeneities of the refractive index described by \eqref{eq:kol_spec} induce phase distortions on the propagating light whose typical coherence length can be quantified (for a plane wave) by the {\it Fried parameter}: $r_0 = (0.42 \, k^2\, C_n^2\, L)^{-3/5}$ \cite{Andrews05}.
Upon propagation, phase distortions combine with diffraction resulting in {\it scintillation}, i.e., turbulence-induced intensity fluctuations, which are generally quantified through the {\it Rytov variance}: $\sigma_R^2 = 1.23 C_n^2 k^{7/6} L^{11/6}$ 
\cite{Andrews05}. Channels with $\sigma_R^2 <1$ are considered to be weakly distorting while those with $\sigma_R^2 >1$ are considered to be strongly distorting.

The refractive index structure constant $C_n^2$ (and consequently $r_0$ and $\sigma_R^2$) varies on typical time-scales of hours \cite{Andrews05} which are much longer than the typical duration of experiments. 
Accordingly, for propagation through horizontal channels (as those considered in this work), $C_n^2$ can be considered to be constant.
Nevertheless, an atmospheric channel evolves in time. According to {\it Taylor's hypothesis} \cite{Taylor35}, for short times ($t \lesssim 10$~ms) \cite{schock2000method}, such evolution can be described as a transverse flow of turbulent eddies due to the presence of wind. 
Under these assumptions, the coherence time of an atmospheric channel is fully defined by the ratio of its coherence length, i.e., the Fried parameter $r_0$, and the average transverse wind velocity $V$ across the channel: $t_c = 0.314\;r_0 / V$ \cite{Roddier04}.

An exact solution to \eqref{eq:stoch_par}, at a fixed time $t$, can be obtained numerically through the {\it split-step} method \cite{Martin88, Smith06, Lukin02}. 
This method consists of dividing the propagation path into segments (each short enough to ensure that intensity fluctuations are small) where all turbulence effects can be described as random phase screens connected by free diffraction in vacuum, with the latter being easily implementable via Fourier optics methods \cite{Goodman05}. 
While at each propagation step only phase distortions are introduced, the combination of refraction on phase screens and diffraction allows to reproduce scintillation effects as well \cite{ Smith06, Lukin02}.
The accuracy of this technique relies on the generation of random phase screens that faithfully obey Kolmogorov statistics, which is possible following a variety of reliable techniques \cite{Martin88, Lane92, Johansson94, Roddier90, bachmann2024accurate, peters2025structured}.   
Within this approach, the temporal evolution of the atmospheric channel can be simulated by transverse shifts of the phase screens according to a Gaussian velocity distribution with mean $V$ and variance $(\Delta V)^2$ \cite{Segel21}.
See supplementary information for further details.

\subsection{The transmission matrix and its real-time estimation}
\label{sec:TMTheory}
The general scheme of our approach in shown in Fig.~\ref{fig:scheme}. Given the linearity of \eqref{eq:stoch_par}, the full propagation properties of an atmospheric channel can be described by a time-dependent transmission matrix $X_t \in \mathbb{C}^{M\times N}$ \cite{Rotter17,Carminati21}.
The matrix $X_t$ maps, at every time $t$, the coefficients $\vec{a}_t \in  \mathbb{C}^{N}$ of a field expansion in a set of transmitter modes $\left\lbrace \phi_j({\bm \rho}, 0)\right\rbrace_{j=0}^{N-1}$, into the coefficients $\vec{y}_t = X_t \vec{a}_t \in  \mathbb{C}^{M}$ of a field expansion in terms of the receiver modes $\left\lbrace \psi_j({\bm \rho}^\prime, L)\right\rbrace_{j=0}^{M-1}$, with ${\bm \rho}$ and ${\bm \rho}^\prime$ the transverse position vectors in the transmitter and receiver planes, respectively.
For an accurate representation of an atmospheric channel, Hermite-Gaussian (HG) modes can be chosen as a convenient transmitter basis $\left\lbrace \phi_j({\bm \rho}, 0)\right\rbrace_{j=0}^{N-1}$ as they are a solution to \eqref{eq:stoch_par} in the absence of turbulence, and therefore feature convenient diffraction properties \cite{Boucher21, Bachmann23}.
On the other hand, at the receiver side, we represent the transmitted light in terms of the computationally and experimentally convenient pixel modes $\left\lbrace \psi_j({\bm \rho}^\prime, L)\right\rbrace_{j=0}^{M-1}$ \cite{Boucher21, Bachmann23}.

Knowledge of the transmission matrix $X_t$ can be used to engineer the optical field $u_t({\bm \rho}, 0)$ in the transmitter aperture, to optimize different properties of the field $u_t({\bm \rho}^\prime, L)$ in the receiver aperture. 
In particular, we will consider a performance metric highly relevant for communication experiments: the single-mode fiber coupling efficiency $\eta(t) = P_f(t)/ P(t)$ where $P_f(t)$ the total power coupled into the fiber and $P(t)$ is the total power available in the plane of the fiber \cite{buck_2004, winzer_1998}
To maximize the coupling efficiency $\eta(t)$, we transmit the {\it phase conjugated} field $u_t({\bm \rho}, 0)$ corresponding to the coefficients $\vec{a}_t = X_t^\dagger \vec{y}_f$, where we introduced the coefficients $\vec{y}_f$ associated with the guided mode of the fiber, back-propagated to the aperture plane of the receiver. See Supplementary information for further details.


To perform the above mentioned optimization, at each time $t$, we need an estimate $\hat{X}_t$ of the transmission matrix $X_t$ of the atmospheric channel. 
We can obtain such estimate using a recursive least-square (RLS) algorithm \cite{Valzania23}, which given the complex inputs $\vec{a}_t$ (prepared with a wavefront shaper) and the complex outputs $\vec{y}_t$ (acquired with a wavefront sensor), and solves the optimization problem $\hat{X}_t = {\rm argmin}_{X_t} \mathcal{L}_t (X_t)$, with
\begin{align}
    \mathcal{L}_t (X_t) = \sum_{\tau = 1}^t \left(\lambda^{t-\tau}||\vec{y}_\tau - X_t \vec{a}_\tau||^2\right) + \delta \lambda^t ||X_t||_F^2
    \label{eq:LS_loss_function}
\end{align}
where $||\cdot||$ and $||\cdot||_F$ are the $L^2$-norm of a vector and the Frobenius norm of a matrix, respectively.
Equation \ref{eq:LS_loss_function} is a linear least square loss function, where, because of the {\it forgetting factor} $0 \ll \lambda < 1$, the {\it data fidelity terms} $||\vec{y}_t - X_t \vec{a}_t||^2$ are weighted exponentially, so that measurements that occurred at $\tau \ll t$ are only marginally relevant to the present estimation.
This feature is crucial to keep up with the temporal evolution of the atmospheric channel.
On the other hand, the Tikhonov regularization term featuring the regularization constant $\delta$, biases the initial estimation of the transmission matrix.
Once $\lambda$ and $\delta$ are fixed, the linear least square problem has a unique solution that can be expressed in the form of normal equation \cite{Valzania23}
\begin{equation}
     C_t \hat{X}_t^\dagger = K_t,
     \label{eq:normal_eq}
\end{equation}
where at each time $t$ the inputs' covariance matrix $C_t$, and input-output cross covariance matrix $K_t$ are defined as
\begin{subequations}
\begin{align}
    C_t &= \sum_{\tau = 1}^t \left(\lambda^{t-\tau}  \vec{a}_\tau \vec{a}_\tau^\dagger \right) + \delta \lambda^t \mathbb{1}_N = \lambda C_{t-1} + \vec{a}_t \vec{a}_t^\dagger, \\
    K_t &= \sum_{\tau = 1}^t \left(\lambda^{t-\tau}  \vec{a}_\tau \vec{y}_\tau^\dagger \right) = \lambda K_{t-1} + \vec{a}_t \vec{y}_t^\dagger,
\end{align}
\label{eq:C_K}
\end{subequations}
with $\mathbb{1}_N$ denoting the $N$-dimensional identity matrix.
It is clear from \eqref{eq:C_K} that, at each time $t$, the least square estimator $\hat{X}_t$ of the transmission matrix can be constructed recursively by adding the new input and output data, $\vec{a}_t$ and $\vec{y}_t$ to the previous estimate of covariance and cross covariance matrices.
Accordingly, the RLS reconstruction algorithm is computationally very appealing, since it takes into account the full history of the channel under investigation, while retaining in memory only information acquired during the previous iteration.
Finally, we point out that while direct inversion of \eqref{eq:normal_eq} is possible, $\hat{X}_t = K_t^\dagger(C_t^{-1})^\dagger$, it is generally preferable to use a much more numerically stable matrix inversion technique, e.g., based on the {\it QR-}decomposition \cite{Alexander93}.

In principle, choosing large transmitter and receiver bases, i.e., large values of $N$ and $M$, allows for a more complete and accurate reconstruction of the transmission matrix $X_t$ \cite{Bachmann23}. 
However, as discussed in the algorithm presented above, the inputs $\vec{a}_t$ are transmitted sequentially. 
In this work we assume realistic bandwidths of both the wavefront shaping and sensing devices. 
With respect to shaping devices, modern micromirror arrays, which can be used to modulate both phase and amplitude, can achieve update rates of $3.6$ kHz at a spatial resolution of $512 \times 320$ pixels \cite{Gehner20}, while binary digital micromirror devices, which in principle could be used to perform the same task at an even higher spatial resolution of $1$ Mpix, can reach update rates of $20$ kHz, albeit with much lower diffraction efficiency (between 1 and 10 \%) \cite{Anzuola16, Yu17}. 
Moreover, interesting phase-only light modulators based on micromirror technology with megapixel counts, $~80\%$ diffraction efficiency and promising $20$ kHz rates have been reported \cite{Rocha:24,Rocha:25}.
As for wavefront sensors, we focus on Shack-Hartmann wavefront sensors which are capable of measurement rates as high as $100$ kHz \cite{Abado10}. 
In this work we consider slightly more modest device capabilities, specifically we assume a bandwidth for both the sensing and shaping devices of $f_{\rm mode} = 5$ kHz.
When we compare this with typical turbulence coherence times for the visible $t_c \sim 1$ ms \cite{Roddier04}, we observe that the relevance of past measurement data diminishes quickly over time.  
Moreover, information from higher-order modes is often less relevant for the optimization of certain performance metrics, such as the fiber coupling efficiency. Higher order modes describe higher spatial frequencies and so will contribute only marginally to overall field. 
To mitigate this effect, it is more convenient to repeatedly transmit a basis with fewer modes, such that the contribution from the most relevant modes is frequently updated, rather than transmit a basis with many (less relevant) modes.
In particular, in our simulations we used $N=10$ HG modes as illustrated in Fig.~\ref{fig:scheme}. 
Accordingly, when the modes are transmitted sequentially, the full $N = 10$ modal basis is transmitted in $N t_{\rm mode} = N/f_{\rm mode} = 2$ ms, where $t_{\rm mode}=0.2$~ms is the time taken to transmit and measure a single mode.
At the receiver side instead we used a square grid with $M = 32\times 32 = 1024$ pixels, comparable with the resolution of standard wavefront sensors \cite{Abado10}.

\subsection{Experimental Setup}

\label{sec:ExpMethods}
\begin{figure*}[t!]
    \centering
    \includegraphics[width = \textwidth]{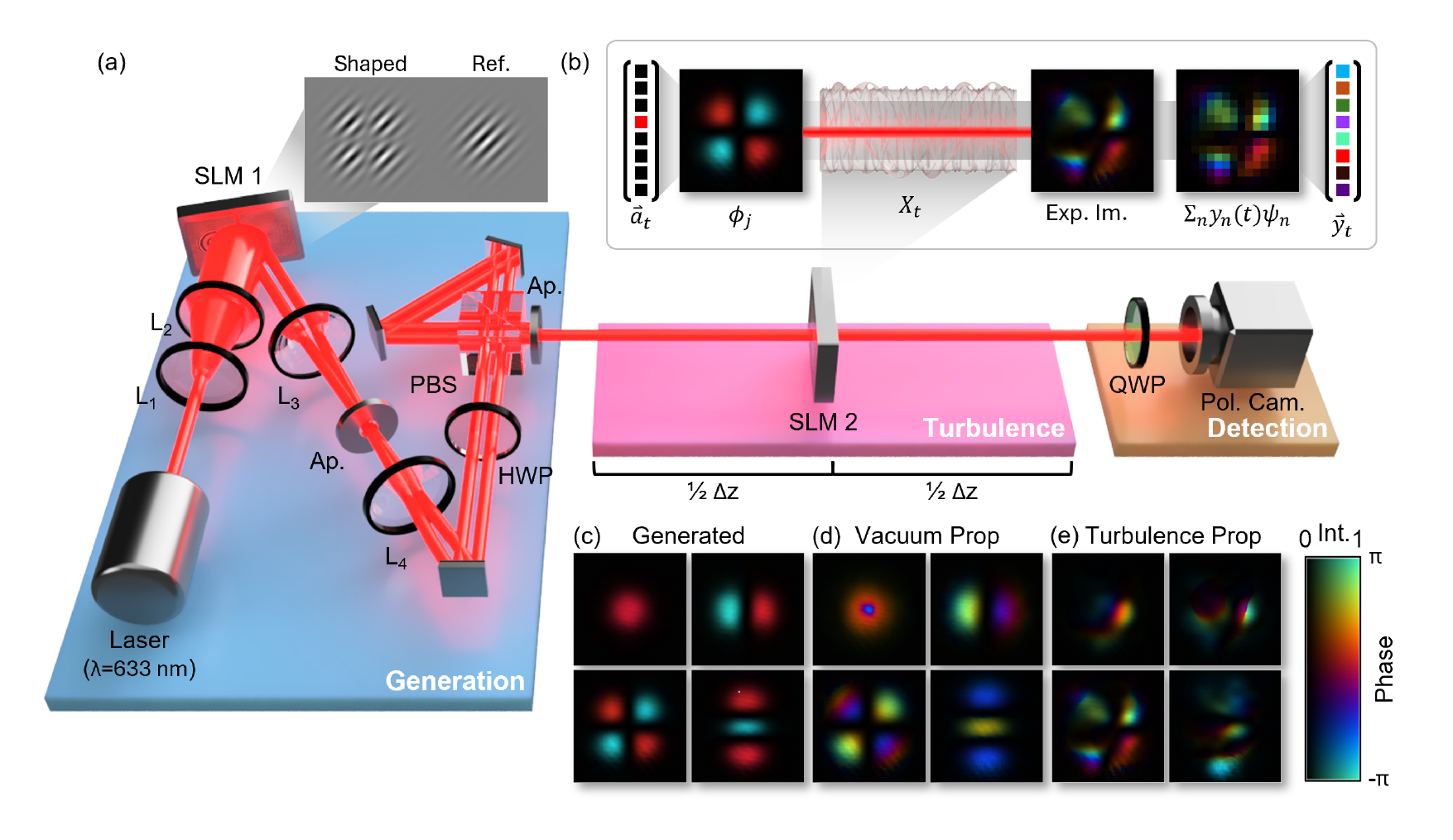}
    \caption{(a) A laser is expanded and collimated onto a spatial light modulator (SLM) which generates a probe beam and a reference beam. These beams are vectorially combined using a Sagnac interferometer before being sent to a second SLM where a phase screen is displayed to simulate the effects of atmospheric turbulence. The full field at the output of the simulated turbulence channel is measured using polarization interferometry. (b) Each input basis vector $\vec{a}_t$ corresponds to a physical HG mode $\phi_j$. The mode propagates through the turbulent channel and exits distorted and has it's amplitude and phase measured by the detector. This information is down-sampled to a resolution of $32\times32$, where each super-pixel represents an output basis mode $\psi_j$. The down-sampled image is a weighted sum of the output basis mode $\sum_n y_n(t)\psi_n$ where the coefficients $y_n(t)$ form the output vector $\vec{y}_t$, representing the transformed input vector $\vec{a}_t$ under the action of the turbulent channel $X_t$. (c) Experimentally measured HG modes before propagation. (d) Experimentally measured HG modes after propagation with no distortion. (e) Experimentally measured HG modes after propagation through simulated turbulence.}
    \label{fig:setup}
\end{figure*}

We test our approach experimentally using the setup shown in Fig.~\ref{fig:setup} (a), which can be conceptually divided into three stages: generation, turbulence, and detection. 
The generation stage consisted of a HeNe laser ($\lambda = 633$~nm) that was expanded and collimated using a $10\times$ objective lens $L_1$ and a plano-convex lens $L_2$ ($f_2=500$~mm), respectively. 
This beam was incident on and overfilled the screen of a spatial light modulator (SLM~1). 
The screen of the SLM was divided into two halves. One half is encoded with the hologram for the beam we wish to transmit through the channel. This could be an HG basis modes $\vec{a}_t$ to probe the channel's TM or shaped beam which has been tailored to optimize a variety of metrics. The other half of the SLM screen is encoded with a Gaussian reference of the same radius $w_0$ as the transmitted HG basis modes. 
The holograms were generated using a complex amplitude modulation scheme allowing us to tailor the fields' amplitude and phase in a single phase element. The hologram was computed using the following expression \cite{arrizon2007pixelated}
\begin{equation}
    H(x,y) = J^{-1}_1 (A(x,y)) \sin[\phi(x,y) + 2\pi(G_x x + G_y y) ] \,,
\end{equation}  
where $A(x,y)$ and $\phi(x,y)$ are the amplitude and phase of the desired field respectively, $J^{-1}_1$ is the inverse Bessel function of the first kind and $G_{x(y)}$ are the grating frequencies in the horizontal and vertical directions. 
Both beams are then sent through a 4f imaging system consisting of lenses $L_3$ and $L_4$ ($f_3=f_4=300$~mm) with a spatial filter in the Fourier plane of $L_3$ to isolate the first diffraction order of the digital hologram. Both the reference and the transmitted HG basis modes or shaped beam were subsequently sent through a half-wave plate (HWP) and Sagnac interferometer where they are vectorially combined  \cite{perez2017demand}. 
At the end of the generation stage, the shaped and  transmitted HG basis modes beam were horizontally polarized and the reference beam was vertically polarized. The beams were then allowed to propagate through our experimentally simulated turbulent channel in order to impart the channel distortion, mathematically described using the transmission matrix $X_t$ for that time step. The channel consists of 0.5~m of free space propagation followed by a second SLM (SLM~2) that is encoded with the desired turbulence phase screen. 
Since the SLM only modulates horizontally polarized light, the phase distortion is only imparted on the shaped beam or the transmitted HG basis modes $\vec{a}_t$, leaving the reference beam unaffected. 
The beams then propagated a further 0.5~m to the end of the channel, where the probe beam or transmitted HG modes have then undergone the full channel transformation, i.e., $\vec{a}_t \rightarrow X_t\vec{a}_t$. 
At the detection stage, a quarter-wave plate (QWP) and polarization sensitive camera (Pol. Cam.) are used to measure the amplitude and phase of the shaped beam and the transmitted HG basis modes using polarization-based interferometry \cite{dudley2014all}. The full field measurements of transmitted HG basis modes were down-sampled into a $32\times32$ super-pixels grid allowing us to decompose the distorted field into the output basis and obtain $\vec{y}_t$. The measured full-field of the shaped beam was used to compute its overlap with the back-propagated fiber mode and calculate the coupling efficiency.
We illustrate the concept of preparing input basis states and measuring output basis states in Fig.~\ref{fig:setup} (b) for use as part of the transmission matrix formalism. Each input basis state is mathematically represented as a vector $\vec{a}_t$, which has a unique entry of 1 and the remaining elements are zero. Each input basis vector $\vec{a}_t$ corresponds to a physical HG mode $\phi_j$, which is prepared by SLM~1. The mode then propagates through the turbulent channel, and becomes distorted due the non-uniform phase profile imparted by SLM~2 in combination with free-space diffraction. At the output of the channel, our detector is able to measure the full field information of the distorted mode. To mimic the resolution of common wavefront sensors, we down-sample this image to a resolution of $32\times32$, where each super-pixel represents an output basis mode $\psi_j$. The down-sampled image is therefore a weighted superposition of these output basis modes, $\sum_n y_n(t)\psi_n$. By using the extracted complex coefficients $y_n(t)$, we create the output vector $\vec{y}_t$, which mathematically represents the transformed input vector $\vec{a}_t$ under the action of the turbulent channel $X_t$.
The measured complex fields of experimentally generated HG modes before and after propagating 1~m are shown in Fig.~\ref{fig:setup} (c) and (d), respectively. Fig.~\ref{fig:setup} (e) shows examples of HG modes after propagating through experimentally simulated turbulence.

\section{Results}
\label{sec:results}

\subsection{Experimental Validation}
\begin{figure*}[t!]
    \centering
    \includegraphics[width = \textwidth]{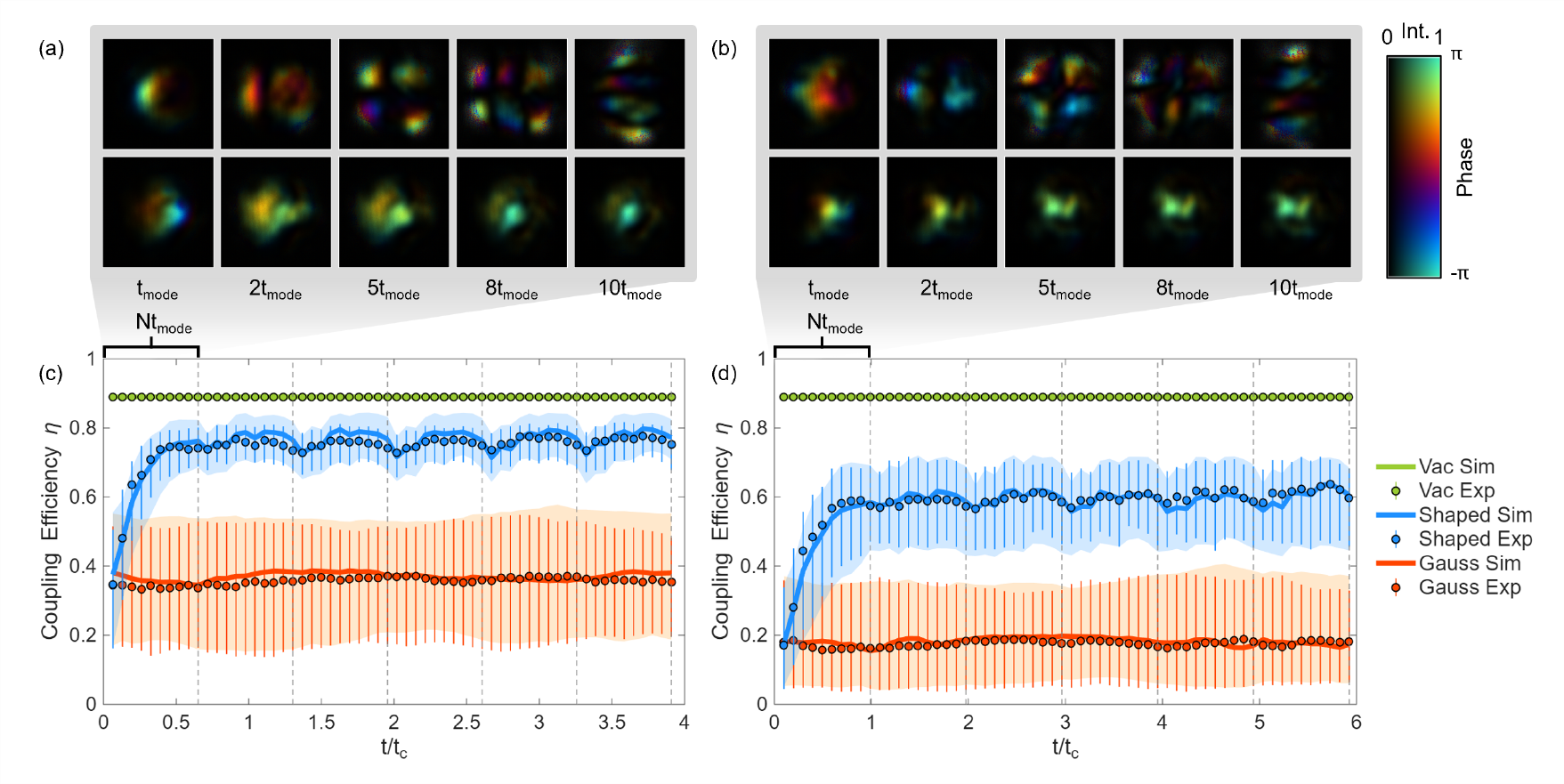}
    \caption{The experimentally measured transmitted HG modes (top row) and the calculated shaped beam for optimal coupling into a fiber at the output of the time-varying turbulent channel (bottom row) for a channel with (a) $\sigma_R^2=0.5$ and (b) $\sigma_R^2=1.0$. The simulated (solid lines) and experimentally measured (points) median coupling efficiency for a Gaussian beam propagating through a vacuum channel (green), unshaped Gaussian beam through a time varying turbulent channel (orange) and a shaped beam computed using our proposed algorithm for channels with (c) $\sigma_R^2=0.5$ and (d) $\sigma_R^2=1.0$.}
    \label{fig:ExpResults}
\end{figure*}

Experimental results are shown in Figure~\ref{fig:ExpResults}, for turbulence strengths quantified by the Rytov variances $\sigma_R^2=0.5$ and $\sigma_R^2=1.0$, where the latter is the strongest turbulence strength that can be reliably simulated using a single phase screen. While the experimental setup consisted of a 1~m long channel, the parameters of the encoded beams, turbulence phase screens and target output field were chosen so that they could be generalized to a real-world link of 1~km using a Fresnel scaling procedure \cite{peters2025structured}. 
We choose channel strengths of $\sigma_R^2 = 0.5$ and $\sigma_R^2 = 1.0$ which correspond to Fried parameters of $r_0 =0.93$~mm and $r_0=0.61$~mm at $633$~nm, respectively. The transverse wind speed was sampled from a distribution with mean $V = 0.095$~m/s and variance $(\Delta V)^2 = 0.032$~m$^2$/s$^2$. Full details of the channel and turbulence phase screens can be found in the Supplementary Information. Figures~\ref{fig:ExpResults}~(a) and (b) show five of the measured HG basis modes after propagating through one realization of each channel during the initial learning period $Nt_{\rm mode}$ (top row) as well as the optimized beam for coupling into the single mode fiber adapted for the time varying channel using the real-time estimated TM (bottom row). As each subsequent HG basis mode is transmitted and uniquely perturbed, the estimation algorithm is able to build up a better estimate of the channel TM. This is evident in the shaped beam, which is scintillated and spread out with a highly non-uniform phase in the initial timestep $t_{\rm mode}$ to centered and focused spot with a constant phase after the first full basis set has been transmitted at $10t_{\rm mode}$. This results in an increase in the fiber coupling efficiency, visually demonstrating the algorithm's effectiveness at overcoming the time-varying distortions imparted by the channel. 

Figures~\ref{fig:ExpResults}~(c) and (d) show the measured coupling efficiency computed from the measured full-field information as function of time for a Gaussian beam with no distortion $\eta_0$ (green), an unshaped Gaussian beam propagating through turbulence $\eta_G$ (orange) and a shaped beam propagating through turbulence optimized using the real-time estimated TM $\eta_{TM}$ (blue). The solid lines indicate the median measured coupling efficiency for numerical simulation matching the exact experimental parameters and the points represent the experimentally measured values. The data was obtained averaging over 200 independent realizations for each channel strength, with the shaded region indicating the inter-quartile range for the simulated data and the error bars indicating the inter-quartile range for the experimental data.  

We first note that there is excellent agreement between simulations of 1~m turbulent channel and the experimental results, with both the median values and inter-quartile ranges matching closely over the full duration of the experiment, with only minor deviations. These simulations were performed in a similar manner to those described in subsequent sections, with the full set of parameters provided in the Supplementary Information.
We observe that at the first timestep, the Gaussian beam and shaped beam have the exact same coupling efficiency at the receiver. 
This is to be expected as the channel has not yet been probed. In the absence of any information about the channel, the optimal choice is to transmit the fundamental Gaussian mode resulting in $\eta_G \approx \eta_{TM}$. However, as the channel begins to evolve in time, we see $\eta_{TM}$ increases up until the end of the first basis repetition $t_0$ (first vertical dotted line). 
This increase can be attributed to the fact that the algorithm is now obtaining more information about the turbulent channel over time by probing it with more HG basis modes. With each subsequent mode transmitted for the first time, the algorithm has arrived at a more accurate estimation of the channel's TM and consequently better optimized the transmitted beam for coupling to the fiber at the output. 
After this point, $\eta_{TM}$ fluctuates periodically around a steady value, changing with a period of $t_0$, in tandem with each new basis repetition. 
Because the oscillation period is equal to $t_0$, we can infer that the information obtained from probing with lower order modes better describes the action of the channel, while higher order modes are not as essential. This is seen by the sudden increase in $\eta_{TM}$ after sending through a Gaussian beam, and the drop off near the end of the basis set.
This behaviour could be avoided by decreasing the basis size which would allow the algorithm to probe the turbulent channel using only the lower order modes and maintain a memory of only the most relevant modes. 
The periodic behaviour is less prominent for $\sigma_R^2=1.0$, which can be understood by realizing that a stronger channel will contain finer features and thus the TM requires higher order modes for an appropriate and accurate estimation. 
In contrast, the median value for $\eta_G$ remains relatively unchanged across the entire measurement period, which is to be expected as no adaptive compensation is being performed to improve its ability to couple into the fibre at the output of the channel. 
We see that after the first basis repetition, $\eta_{TM}$ remains consistently higher for both turbulence strengths, exhibiting a $\approx40\%$ improvement over the measurement period. 
In addition to the median improvement, we can also see that the variation in the coupling efficiency is much smaller for $\eta_{TM}$ in comparison to $\eta_G$ in both channels. 
We therefore observe that estimating the TM does not only increase the median performance of the link, but also increases the consistency in the performance, making any information transmission through  the complex, time-varying channel far more reliable.

\subsection{Deep Turbulence Simulations}

\begin{figure*}[ht!]
    \centering
    \includegraphics[width = \textwidth]{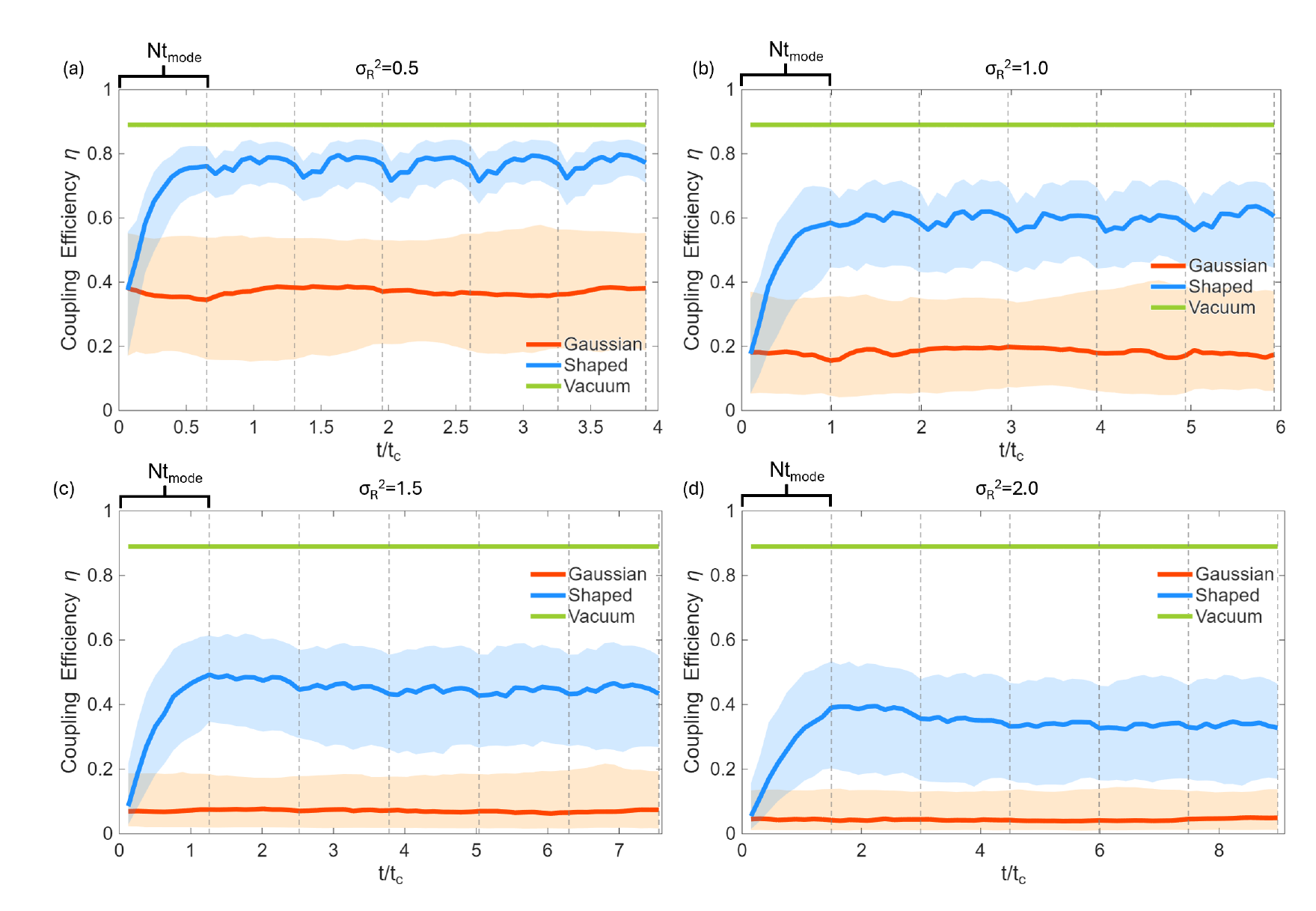}
    \caption{ The median coupling efficiency for a Gaussian beam propagating through a vacuum channel (green), unshaped Gaussian beam sent through a time varying turbulent channel (orange) and a shaped beam computed using our proposed algorithm for simulated 1~km channels with (a) $\sigma_R^2=0.5$, (b) $\sigma_R^2=1.0$, (b) $\sigma_R^2=1.5$ and (d) $\sigma_R^2=2.0$. The median was taken over 500 independent realizations with the shaded region showing the inter-quartile range.}
    \label{fig:SimResults}
\end{figure*}

We have now demonstrated that our real-time estimation approach is effective at compensating for turbulence both in simulation and experiment using a tabletop turbulence simulator. Given the almost perfect agreement between simulation and experiment demonstrated in those results, we now extend our approach and demonstrate its efficacy in a $L=1$~km long atmospheric using numerical simulations. This simulated channel is delimited by transmitter and receiver apertures of equal diameter $D = 3.79$~cm, for a monochromatic beam of wavelength $\lambda  = 633$~nm. We study channels in the weak ($\sigma_R^2$<1), moderate ($\sigma_R^2\approx1$) and strong ($\sigma_R^2>1$) regimes and again opted to leverage the estimated TM to optimize the power coupling efficiency $\eta$ of the transmitted beam into a single-mode fiber. 

We show the results in Figure~\ref{fig:SimResults} (a)-(d) for turbulence strengths $\sigma_R^2 = 0.5,\,1.0,\,1.5$ and $2.0$ respectively. The green line indicates the coupling efficiency of a Gaussian beam with a waist $w_0 = 16$~mm at the receiver after propagating through a vacuum channel, achieving a constant $\eta_0 \approx89\%$ in the absence of any distortion. We show the median coupling efficiency of the same Gaussian beam through simulated turbulence ($\eta_G$) in orange  and the coupling efficiency of the shaped beam ($\eta_{TM}$) in blue , computed using the real-time estimated TM of the channel. The results are computed over 500 independent realizations for each channel with the shaded regions indicating the inter-quartile range over these measurements. The dotted vertical lines indicate a new basis repetition for the probing HG modes. The channels had refractive index structure constants of $C_n^2= 8.83 \times10^{-15},\,1.77 \times10^{-14},\,2.65 \times10^{-14}$ and $3.53 \times10^{-14}$~m$^{-2/3}$ which corresponded to Fried parameters for the channel of $r_0 = 2.9,\,1.9,\,1.5$ and $1.2$~cm, respectively. Each simulation used four evenly spaced phase screens, with the Fried parameters of $r_{0,s} = 6.6,\,4.9,\,3.4$ and $2.9$~cm respectively for each channel strength. Furthermore, we assumed the transverse wind speed to have a Gaussian distribution with mean $V = 3$~m/s and variance $(\Delta V)^2 = 1\;{\rm m}^2/{\rm s}^2$, which results in the channels having coherence times $t_c = 6.94,\,4.58,\,3.59$ and $3.02$~ms, respectively for each turbulence strength at a wavelength of $633$~nm. 

Figures~\ref{fig:SimResults} (a) and (b) agree very well with the bench-top experimental results as shown in Figures~\ref{fig:ExpResults} (c) and (d), exhibiting similar coupling efficiencies for the vacuum, Gaussian, and shaped beams. This is to be expected, as the parameters chosen for the experimental investigation were chosen to scale faithfully to channels closer to real-world distances and conditions. The explanations and discussions of trends and behaviours seen in Figures~\ref{fig:ExpResults} (c) and (d) are therefore equally applicable to the simulated results shown in Figures~\ref{fig:SimResults} (a) and (b).
Curiously, we are able to see 3 distinct behaviors of the algorithm once it has fully probed the channel, dictated by the relationship between $Nt_{\rm mode}$ and the channel coherence time $t_c$. In Figures~\ref{fig:SimResults}~(a) and (b), we have weak to moderate turbulent channels with $Nt_{\rm mode}\leq t_c$. In these regimes, we observe periodic behavior in the algorithm's performance that coincides with the start of a new HG basis repetition. Here, the algorithm probes the channel fast enough to obtain sufficient information to appropriately shape the transmitted beam, leading to a steady average over long timescales. The periodic nature is due to the weak transverse phase distortion, dominated by low order modes with the effects of highest order modes being negligible and whose probing caused the estimated TM to be out of date, leading to the slight drop in $\eta_{TM}$. This is confirmed when the basis repeats again. The information of more recently transmitted lower order modes is included which results in a sudden increase in coupling efficiency. We do observe that the performance decreases as the channel strength increases and that periodic nature is less prominent in the stronger of the two channels, likely due to the channel evolving more rapidly. Figures~\ref{fig:SimResults}~(c) and (d) show channels in the strong fluctuation regime where $Nt_{\rm mode}>t_c$ and we observe no periodic features in the algorithm's performance. We can attribute this to two factors. First, the stronger phase distortions in the channel now require the finer detailed features from higher order HG modes to be sufficiently compensated for. Second, the channel is now sampling a full basis mode set while the channel distortions are significantly changing, making it harder to arrive at an optimal solution. Consequently, we see that the performance remains approximately constant after the initial learning period $Nt_{\rm mode}$, with a slight but steady drop as the channel evolves. 

\begin{figure*}[ht!]
    \centering
    \includegraphics[width = \textwidth]{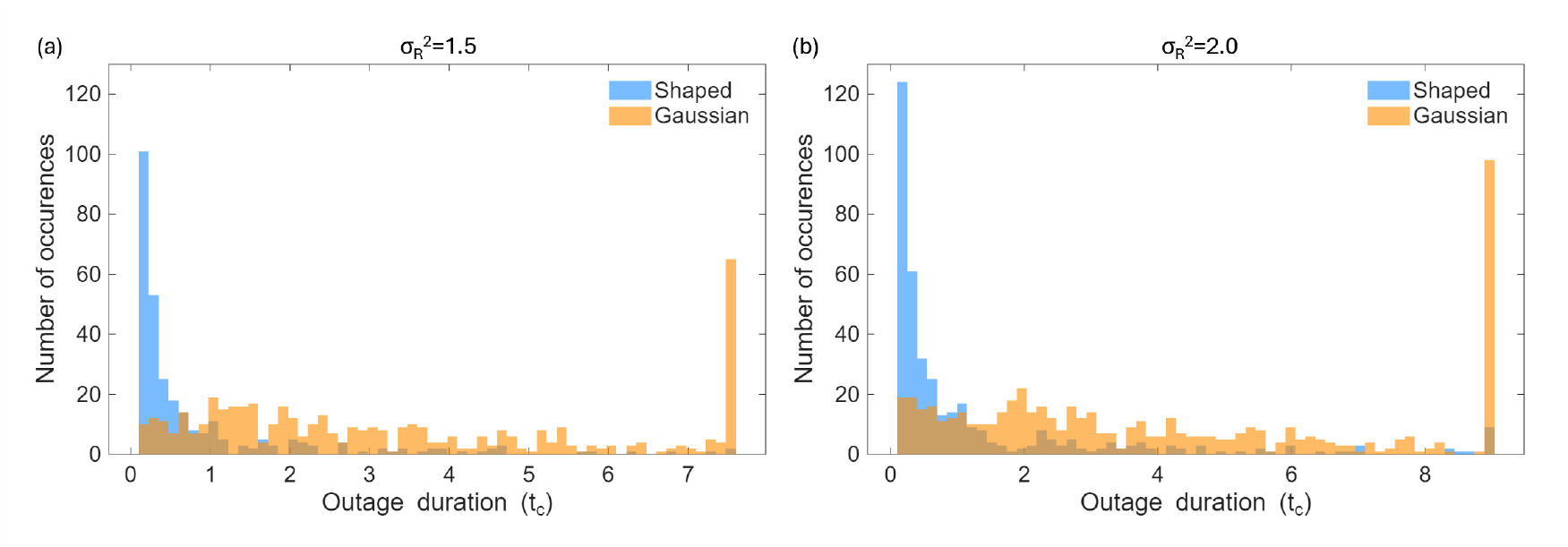}
    \caption{The number of outages versus the outage duration for a Gaussian beam and the shaped beam for 500 independent realizations of a time-varying turbulence channel with (a) $\sigma_R^2=0.5$ and (b) $\sigma_R^2=1.0$}
    \label{fig:Outages}
\end{figure*}

An additional observation can be made from Figures~\ref{fig:SimResults}~(a) and (b). In both cases, the inter-quartile range of the shaped beam is noticeably smaller than that of the unshaped beam, indicating that the coupling efficiency of the shaped beam was far more consistent as the medium changed over time. This matches the observation in the bench-top experiment and indicates that our method of estimating the TM in real-time does not only increase the median performance of the link, but also makes the performance more reliable over time. This increased consistency cannot be directly seen in the results in Figures~\ref{fig:SimResults}~(c) and (d) as these are in the strong fluctuation regime. Here, the coupling efficiency of the unshaped beam is already very poor ($\approx5\%$), resulting in little variation around this very low value as compared to the weaker turbulence strengths in Figures~\ref{fig:SimResults}~(a) and (b). However, the increase in reliability can be seen in this regime if we look at the number and duration of outage events at the receiver. We define an outage event to occur when the coupling efficiency at the receiver drops below 5\% and show these results in Figures~\ref{fig:Outages} (a) and (b) for $\sigma_R^2 = 1.5$ and $\sigma_R^2 = 2.0$, respectively. First, we observe that shaping the input beam using the estimated TM decreases the total number of outages experienced by the system, dropping from 466 to 296 for $\sigma_R^2 = 1.5$ and from 567 to 392 for $\sigma_R^2 = 2.0$. Second, we find that if an outage event does occur, its duration for the shaped beam is significantly shorter than that for the unshaped Gaussian beam. The large spike at the end of the plot counts any outages that last longer than the entire measurement period. As we can see, there are a significant number of realizations where this occurs for the Gaussian beam, and only happens a handful of times for the shaped beam. The outages in the case of the shaped beam primarily occur during the initial learning period $Nt_{\rm mode}$, as can be seen in Figure S1 of the Supplementary Information. This is confirmed by looking at both Figures~\ref{fig:Outages} (a) and (b), with most of the shaped outage events being shorter than $t_c$, while the unshaped outage events had a uniform distribution of durations and a large spike near the longest duration measured. If we exclude the initial learning period, the number of outage events drop from 417 to 104 for $\sigma_R^2 = 1.5$ and from 512 to 160 for $\sigma_R^2=2.0$. With this context we observe that the implementation of our approach reduces the number of outages by $\approx 70\%$. 

\section{Conclusion}


In this work, we have proposed a method for acquiring the transmission matrix of an atmospheric channel using a recursive real-time estimation technique and tested it under weak, moderate, and strong turbulence conditions. 
We have shown its accuracy and utility by optimizing a communication-relevant performance metric of paramount importance: the coupling efficiency into a single-mode fiber $\eta(t)$, which is vital for coherent optical communication schemes but highly sensitive to the fine details in the field distribution at the receiver. 
Without the use of adaptive compensation, we observed that $\eta$ is severely degraded after propagation through a time-varying turbulence channel, a result of the phase distortions and scintillation induced by the medium's perturbations. For real-world communications, this leads to a significant increase in bit-error and a decrease in information capacity \cite{fried1967optical,dikmelik2005fiber}. However, using our approach and estimating, even imperfectly, the TM of the atmospheric channel, has a significant impact on the optimization of the fiber coupling efficiency $\eta(t)$, increasing it by 30\%-40\%, increasing its stability in weak turbulent conditions, and greatly reducing the number and duration of outage events in strong turbulence conditions. Throughout our investigation, we have matched our simulation and experimental parameters with that of state-of-the-art structured mode generators and wave-front sensors to ensure real-world applicability of the technique.


Finally, in the present work we focused on horizontal channels, and on communication-relevant metrics.
However, our approach can be easily adapted to consider slant or vertical paths and the estimated transmission matrix can be used for several different tasks such as focusing \cite{Vellekoop07, Popoff10} and point-spread function engineering \cite{Boniface17}.
We made use of a polarization based wavefront sensor in this work, with interference captured with a CMOS camera. Real-world implementation of our approach would leverage more sophisticated and high-speed technologies, with current state-of-the-art in the development of wavefront sensors being able to reliably operate at high-speed and in strong turbulence conditions \cite{Zepp13,Zepp21,Zepp22}. Our findings pave the way to significantly enhance free-space communication and remote imaging both in ground-to-ground and ground-to-space applications.

\begin{backmatter}
\bmsection{Funding} This work was carried out during the tenure of an ERCIM ‘Alain Bensoussan’ Fellowship Programme.
This work was supported by the Fraunhofer Internal Programs under Grant No. Attract 40-09467.

\bmsection{Acknowledgments} GS is grateful to S. Gigan (LKB) and L. Valzania (Greenerwave) for their insights on the real-time estimation of the transmission matrix, and D. Bachman (Uni. Freiburg) for our discussion on wave optics simulations of atmospheric channels.

\bmsection{Disclosures} The authors declare no conflicts of interest.

\bmsection{Data Availability Statement} Data is available from the corresponding author upon reasonable request.

\bmsection{Supplemental document} See supplemental document for supporting content.

\end{backmatter}

\newpage
\clearpage
\appendix
\onecolumn

\setcounter{section}{0}
\setcounter{figure}{0}
\setcounter{table}{0}
\setcounter{equation}{0}
\setcounter{footnote}{0}
\renewcommand{\thesection}{SUPPLEMENTARY \arabic{section}}
\renewcommand{\thefigure}{S\arabic{figure}}
\renewcommand{\thetable}{S\arabic{table}}
\renewcommand{\theequation}{S\arabic{equation}}


\section{Parameters for experimental demonstration}
\label{sec:wave-optics}
Parameters of the channel geometry used in the experimental demonstration are shown in Table~\ref{tab:ExpParams} and the parameters for the turbulent phase screens are shown in Table~\ref{tab:ExpTurb}. The experiment was performed using a 1~m long channel and implemented only a single phase screen to impart the phase perturbations. The use of a single phase screen allowed for testing our approach in the weak ($\sigma_R^2$<1) and moderate ($\sigma_R^2\approx1$) turbulent regimes. 

\begin{table*}[h!]
\centering
\caption{Parameters of the channel geometry for the experimental demonstration.}
\begin{tabular}{ccc}
{\bf Geometrical parameters} &  \\
         \hline
         Input aperture  & \multicolumn{2}{c}{$D_{\rm in} = D = 2.07$ mm} \\
         Output aperture  & \multicolumn{2}{c}{$D_{\rm out} = D = 2.07$ mm} \\
         Channel length & \multicolumn{2}{c}{$L = 1$~m} \\
         {\bf Mode parameters} &   \\
         \hline
         Input HG modes' waist & \multicolumn{2}{c}{$w_0 = 0.5$ mm} \\
         Number of input modes & \multicolumn{2}{c}{HG modes with $n+m \leq 3$ ($N=10$ modes in total)} \\
         Output pixel modes size & \multicolumn{2}{c}{$d_{out} = 64.7  $~\textmu m} \\
         Output pixels number & \multicolumn{2}{c}{$M = 32 \times 32 = 1024$} \\
         {\bf Fiber coupling} &   \\
         \hline
         Fiber Type & \multicolumn{2}{c}{Single-Mode: S630-HP \cite{Thorlabs_2024}} \\
         Optical wavelength & \multicolumn{2}{c}{$\lambda = 633$ nm} \\
         Optimal mode waist (focal) & \multicolumn{2}{c}{$w_{\rm target} = 4.12$ \textmu m} \\
         Focal length & \multicolumn{2}{c}{$f = 10$~mm} \\
\end{tabular}
  \label{tab:ExpParams}
\end{table*}

\begin{table*}[h!]
\centering
\caption{Parameters used to simulate turbulence in the experimental demonstration.}
\begin{tabular}{ccc}
         {\bf Turbulence parameter} & Weak & Moderate  \\
         \hline
         Rytov variance $\sigma_R^2$ & $0.5$ & $1.0$   \\
         Structure constant $C_n^2$ (${\rm m}^{-2/3}$) & $2.8\times 10^{-9}$ & $5.6\times 10^{-9}$ \\
         Channel Fried parameter $r_0$ (mm) & $0.93$ & $0.61$ \\
         Coherence time $t_c$ (ms) & $3.07$ & $2.03$ \\
         Mean wind speed $\bar{V}$ (m/s) & \multicolumn{2}{c}{0.095}\\
         Wind speed variance $\Delta V$ (m$^2$/s$^2$) & \multicolumn{2}{c}{0.032}
\end{tabular}
  \label{tab:ExpTurb}
\end{table*}

\section{Wave-optics simulations of atmospheric channels}
\label{sec:wave-optics}
We performed wave optics simulations on a $1024 \times 1024-$pixels numerical grid with the size of four times the diameter, $D = 6.55$~cm, of the transmitter and receiver apertures. 
To simulate free space propagation in vacuum, we used the angular spectrum propagator \cite{Goodman05, Schmidt10}. We generated random phase screens with Kolmogorov statistics (i.e. we did not include any inner or outer scales except for those enforced by the pixel and grid sizes in our numerical simulations) using the algorithm from \cite{Lane92} with $5$ subharmonics levels. 
To ensure the accuracy of our numerical simulations, we enforced a Rytov variance $\sigma_R^2 < 1$ for each partial propagation step. 
Accordingly, 4 screens were sufficient to simulate a wide range of turbulent channels with Rytov variances ranging from $\sigma_R^2=0.5$ to $\sigma_R^2=2.0$. 
All parameters of our wave optics simulations are summarized in Tables~\ref{tab:SimParams} and \ref{tab:SimTurb}.

\begin{table*}[h!]
\centering
\caption{Parameters of the channel geometry for the wave-optics simulations.}
\begin{tabular}{ccc}
{\bf Geometrical parameters} &  \\
         \hline
         Input aperture  & \multicolumn{2}{c}{$D_{\rm in} = D = 6.55$~cm} \\
         Output aperture  & \multicolumn{2}{c}{$D_{\rm out} = D = 6.55$~cm} \\
         Channel length & \multicolumn{2}{c}{$L = 1$~km} \\
         {\bf Mode parameters} &   \\
         \hline
         Input HG modes' waist & \multicolumn{2}{c}{$w_0 = 15.8$~mm} \\
         Number of input modes & \multicolumn{2}{c}{HG modes with $n+m \leq 3$ ($N=10$ modes in total)} \\
         Output pixel modes size & \multicolumn{2}{c}{$d_{out} = 2.05$ mm} \\
         Output pixels number & \multicolumn{2}{c}{$M = 32 \times 32 = 1024$} \\
         {\bf Fiber coupling} &   \\
         \hline
         Fiber Type & \multicolumn{2}{c}{Single-Mode: S630-HP \cite{Thorlabs_2024}} \\
         Optical wavelength & \multicolumn{2}{c}{$\lambda = 633$ nm} \\
         Optimal mode waist (focal) & \multicolumn{2}{c}{$w_{\rm target} = 4.12$ \textmu m} \\
         Focal length & \multicolumn{2}{c}{$f = 0.32$ m} \\
\end{tabular}
  \label{tab:SimParams}
\end{table*}

\begin{table*}[h!]
\centering
\caption{Parameters used to simulate turbulence in the wave-optics simulations.}
\begin{tabular}{ccccc}
         {\bf Turbulence parameter} & Weak & Moderate & \multicolumn{2}{c}{Strong}  \\
         \hline
         Rytov variance $\sigma_R^2$ & $0.5$ & $1.0$ & $1.5$ & $2.0$  \\
         Structure constant $C_n^2$ (${\rm m}^{-2/3}$) & $8.83\times 10^{-15}$ & $1.77\times10^{-14}$ & $2.65\times 10^{-14}$ & $3.53\times10^{-14}$\\
         Step Fried parameter $r_{0,s}$ (mm) & $66.3$ & $43.8$ & $34.3$ & $28.9$ \\
         Channel Fried parameter $r_0$ (mm) & $28.9$ & $19.0$ & $14.9$ & $12.6$ \\
         Coherence time $t_c$ (ms) & $3.02$ & $1.99$ & $1.56$ & $1.32$\\
         Mean wind speed $\bar{V}$ (m/s) & \multicolumn{4}{c}{3.0}\\
         Wind speed variance $\Delta V$ (m$^2$/s$^2$) & \multicolumn{4}{c}{1.0}
\end{tabular}
  \label{tab:SimTurb}
\end{table*}

\section{Calculation of Fiber Coupling Efficiency}
\label{sec:fiber}

The degree to which an arbitrary signal can be coupled to an optical fiber is quantified by the coupling efficiency $\eta$, which is defined as the ratio of optical power coupled into the fiber $P_{c}$ to the available incident optical power in the plane of the fiber $P_{in}$. Ignoring losses due to Fresnel reflection at the fiber facet, the coupling efficiency for a single-mode fiber is given by the overlap integral \cite{buck_2004}
\begin{equation}
        \eta = \frac{P_{c}}{P_{in}} = \frac{\left| \iint_S u_{{in}}({\bm \rho}) \;  u^{*}_{0}({\bm \rho}) \, dS \right|^2}{ \iint_S \left| u_{{in}}({\bm \rho}) \right|^2 \, dS \; \iint_S \left| u_{0}({\bm \rho}) \right|^2 \, dS},
    \label{eq:coupling_efficiency}
\end{equation}
where $u_{in}({\bm \rho})$ and $u_{0}({\bm \rho})$ are the complex fields of the signal incident on the fiber and the guided mode of the fiber respectively, ${\bm \rho}$ is a transverse position vector located in the plane of the fiber, $^*$ denotes complex conjugation, and the integrals are evaluated over the entire (infinite) transverse plane at the fiber facet $S$. 

For a step-index optical fiber with a small refractive index constant (i.e.: weakly-guiding), the guided mode can be well described using the linearly polarized (LP) mode approximation and expressed as \cite{buck_2004, agrawal_2010}:
\begin{equation}
    u_0({\bm \rho}) = A_0
    \begin{cases} 
      J_0\left( \frac{p}{a} \left| \bm \rho \right| \right) & \left| \bm \rho \right| \leq a \\
      \frac{J_0\left( p \right)}{K_0\left( q \right)} K_0\left( \frac{q}{a} \left| \bm \rho \right| \right) & \left| \bm \rho \right| > a
    \end{cases}
    \label{eq:guided_mode}
\end{equation}
where $A_0$ is a scaling constant related to the power in the mode, $J_0\left( \cdot \right)$ is the Bessel function of the first kind, $K_0\left( \cdot \right)$ is the modified Bessel function of the second kind, $a$ is the radius of the fiber core, and $p$ and $q$ are dimensionless parameters related to the design of the fiber. In practice, the values of $p$ and $q$ are determined by numerically solving the fiber eigenvalue equation, given the core radius $a$, the fiber's numerical aperture $\textnormal{NA}$, and the wavelength of the signal $\lambda$. 

To compute the coupling efficiency for fields propagated through the atmospheric channels described in Sec.~\ref{sec:wave-optics}, we consider a simple optical system consisting of a single lens in the aperture plane which focuses the received beam to the fiber facet. The focal length of the lens used is chosen to optimize the coupling of the received Gaussian beam in the absence of turbulence. Specifically, a lens with focal length $f = 0.35$ cm is chosen such that the waist of the focused beam matches the waist of the guided fiber mode $\omega_0 = $ 4.61 \textmu m (which we estimate by approximating the fiber mode as a Gaussian beam \cite{marcuse_1977}) for a standard Single-Mode S630-HP \cite{Thorlabs_2024} at $\lambda =$ 633~nm. The same fiber is considered to numerically evaluate the expression of the guided mode given in Equation \ref{eq:guided_mode}. To avoid the computationally expensive operation of propagating each field from the aperture plane to the plane of the fiber, we rather back-propagate the guided mode through the coupling lens, which allows the integrals in Equation \ref{eq:coupling_efficiency} to be evaluated in the aperture plane \cite{winzer_1998}. The coupling efficiency is then computed numerically using the fields received in the aperture plane obtained via the wave optics simulations described in Sec.~\ref{sec:wave-optics}.

\section{Outage Plots}
\label{sec:outage}

\begin{figure*}[ht!]
    \centering
    \includegraphics[width = \textwidth]{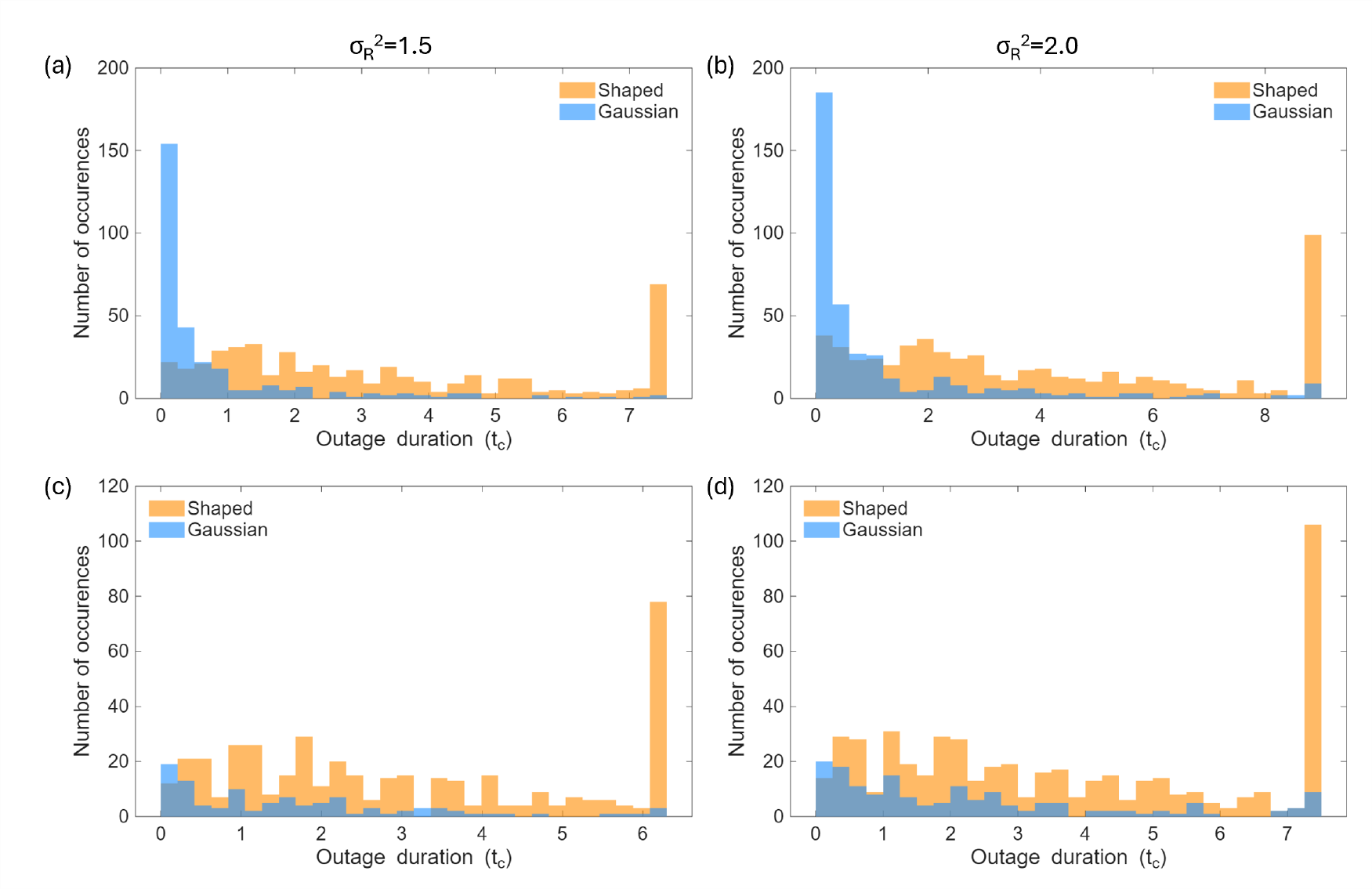}
    \caption{The number of outages versus the outage duration for a Gaussian beam and the shaped beam for 500 independent realizations of a time varying turbulence channel with (a) $\sigma_R^2=1.5$ and (b) $\sigma_R^2=2.0$ and with outages during the initial learning period $Nt_{\rm mode}$ for (c) $\sigma_R^2=1.5$ and (d) $\sigma_R^2=2.0$}
    \label{fig:OutagesSupp}
\end{figure*}

\end{document}